\documentclass[conference,letterpaper]{IEEEtran}
\usepackage{graphicx}
\usepackage{subfigure,stfloats}
\usepackage{amssymb}
\usepackage{tikz}

\usepackage[utf8]{inputenc} 
\usepackage[T1]{fontenc}
\usepackage{url}
\usepackage{ifthen}
\usepackage{cite}
\usepackage[cmex10]{amsmath} % Use the [cmex10] option to ensure complicance
\newtheorem{theorem}{Theorem}
\newtheorem{lemma}{Lemma}

\newtheorem{proposition}{Proposition}
\newtheorem{remark}{Remark}

\newcommand\phZ[1][{}]{\vphantom{$0$}\makebox[0pt][c]{\small\boldmath$#1$}\vphantom{$0$}}
\def\Ii{i}
\def\Iii{\mbox{\bf-}\!1}
\def\Iiii{\mbox{\bf-}\hspace{-0.10em}i}
\def\Iiiii{1}
\def\Aa#1#2#3#4{#1}
\def\Bb#1#2#3#4{#2}
\def\Cc#1#2#3#4{#3}
\def\Dd#1#2#3#4{#4}
\def\pcSH#1#2#3#4#5{\pcSh{#2}{#3}{#4}{#5}{#1}}

\def\pcSh#1#2#3#4#5{
\begin{tikzpicture}[
xscale=0.43,yscale=0.36,xslant=-0.5,
nn/.style={circle,fill=#5,draw=black,%thick, 
           inner sep=1pt}]
           \begin{scope}  
\clip [] (-0.45,-0.45) rectangle (3.45,3.45);
\draw[ystep=1,xstep=1,] (-4.9,-2.1) grid (5.4,3.9);
\draw[xslant=1,ystep=9,xstep=1,] (-3.4,-2.1) grid (6.4,3.9);
\end{scope}
\draw 
  (0,0) node [nn] {\phZ[\Aa#1]} +(1,0) node [nn] {\phZ[\Bb#1]} +(2,0) node [nn] {\phZ[\Cc#1]} +(3,0) node [nn] {\phZ[\Dd#1]}
++(0,1) node [nn] {\phZ[\Aa#2]} +(1,0) node [nn] {\phZ[\Bb#2]} +(2,0) node [nn] {\phZ[\Cc#2]} +(3,0) node [nn] {\phZ[\Dd#2]}
++(0,1) node [nn] {\phZ[\Aa#3]} +(1,0) node [nn] {\phZ[\Bb#3]} +(2,0) node [nn] {\phZ[\Cc#3]} +(3,0) node [nn] {\phZ[\Dd#3]}
++(0,1) node [nn] {\phZ[\Aa#4]} +(1,0) node [nn] {\phZ[\Bb#4]} +(2,0) node [nn] {\phZ[\Cc#4]} +(3,0) node [nn] {\phZ[\Dd#4]};
\end{tikzpicture}
}

\def\picEEEE{
\makebox[0mm][l]{$\chi_{-\omega}$:}\!\!\!
\pcSH{white}{{\Iiiii}{\Ii}{\Iii}{\Iiii}} {{\Ii}{\Iii}{\Iiii}{\Iiiii}} {{\Iii}{\Iiii}{\Iiiii}{\Ii}} {{\Iiii}{\Iiiii}{\Ii}{\Iii}} \nolinebreak
\ \ \makebox[0mm][l]{$\chi_{-\psi}$:}\!\!\!
\pcSH{gray!40!white}{{\Iiiii}{\Iiii}{\Iii}{\Ii}} {{\Iiiii}{\Iiii}{\Iii}{\Ii}} {{\Iiiii}{\Iiii}{\Iii}{\Ii}} {{\Iiiii}{\Iiii}{\Iii}{\Ii}}\nolinebreak
\ \ \makebox[0mm][l]{$\chi_{-\bar\omega\psi}$:}\!\!\!
\pcSH{gray!40!white}{{\Iiiii}{\Ii}{\Iii}{\Iiii}} {{\Iiii}{\Iiiii}{\Ii}{\Iii}} {{\Iii}{\Iiii}{\Iiiii}{\Ii}} {{\Ii}{\Iii}{\Iiii}{\Iiiii}} \nolinebreak
\ \ \makebox[0mm][l]{$\chi_{\bar\omega}$:}\!\!\!
\pcSH{white}{{\Iiiii}{\Iiii}{\Iii}{\Ii}} {{\Iii}{\Ii}{\Iiiii}{\Iiii}} {{\Iiiii}{\Iiii}{\Iii}{\Ii}} {{\Iii}{\Ii}{\Iiiii}{\Iiii}}
\\[-1mm]

\mbox{}\hspace{18mm}
\makebox[0mm][l]{$\chi_{2\omega}$:}\!\!\!
\pcSH{white}{{\Iiiii}{\Iii}{\Iiiii}{\Iii}}{{\Iii}{\Iiiii}{\Iii}{\Iiiii}}{{\Iiiii}{\Iii}{\Iiiii}{\Iii}}{{\Iii}{\Iiiii}{\Iii}{\Iiiii}}\nolinebreak
\ \ \makebox[0mm][l]{$\chi_{-\omega\psi}$:}\!\!\!
\pcSH{gray!40!white}{{\Iiiii}{\Iiiii}{\Iiiii}{\Iiiii}}{{\Ii}{\Ii}{\Ii}{\Ii}}{{\Iii}{\Iii}{\Iii}{\Iii}}{{\Iiii}{\Iiii}{\Iiii}{\Iiii}}\nolinebreak
\ \ \makebox[0mm][l]{$\chi_{2\bar\omega}$:}\!\!\!
\pcSH{white}{{\Iiiii}{\Iii}{\Iiiii}{\Iii}}{{\Iiiii}{\Iii}{\Iiiii}{\Iii}}{{\Iiiii}{\Iii}{\Iiiii}{\Iii}}{{\Iiiii}{\Iii}{\Iiiii}{\Iii}}\nolinebreak
\ \ \makebox[0mm][l]{$\chi_{\omega\psi}$:}\!\!\!
\pcSH{gray!40!white}{{\Iiiii}{\Iiiii}{\Iiiii}{\Iiiii}}{{\Iiii}{\Iiii}{\Iiii}{\Iiii}}{{\Iii}{\Iii}{\Iii}{\Iii}}{{\Ii}{\Ii}{\Ii}{\Ii}}
\\[-1mm]

\mbox{}\hspace{18mm}\mbox{}\hspace{18mm}
\makebox[0mm][l]{$\chi_\omega$:}\!\!\!
\pcSH{white}{{\Iiiii}{\Iiii}{\Iii}{\Ii}}{{\Iiii}{\Iii}{\Ii}{\Iiiii}}{{\Iii}{\Ii}{\Iiiii}{\Iiii}}{{\Ii}{\Iiiii}{\Iiii}{\Iii}}\nolinebreak
\ \ \makebox[0mm][l]{$\chi_{-\bar\omega}$:}\!\!\!
\pcSH{white}{{\Iiiii}{\Ii}{\Iii}{\Iiii}}{{\Iii}{\Iiii}{\Iiiii}{\Ii}}{{\Iiiii}{\Ii}{\Iii}{\Iiii}}{{\Iii}{\Iiii}{\Iiiii}{\Ii}}\nolinebreak
\ \ \makebox[0mm][l]{$\chi_{\bar\omega\psi}$:}\!\!\!
\pcSH{gray!40!white}{{\Iiiii}{\Iiii}{\Iii}{\Ii}}{{\Ii}{\Iiiii}{\Iiii}{\Iii}}{{\Iii}{\Ii}{\Iiiii}{\Iiii}}{{\Iiii}{\Iii}{\Ii}{\Iiiii}}\nolinebreak
\ \ \makebox[0mm][l]{$\chi_{\psi}$:}\!\!\!
\pcSH{gray!40!white}{{\Iiiii}{\Ii}{\Iii}{\Iiii}}{{\Iiiii}{\Ii}{\Iii}{\Iiii}}{{\Iiiii}{\Ii}{\Iii}{\Iiii}}{{\Iiiii}{\Ii}{\Iii}{\Iiii}}
\\[-1mm]

\mbox{}\hspace{18mm}\mbox{}\hspace{18mm}\mbox{}\hspace{18mm}
\makebox[0mm][l]{$\chi_0$:}\!\!\!
\pcSH{gray}{{1}{1}{1}{1}}{{1}{1}{1}{1}}{{1}{1}{1}{1}}{{1}{1}{1}{1}}\nolinebreak
\ \ \makebox[0mm][l]{$\chi_1$:}\!\!\!
\pcSH{white}{{1}{\Iii}{1}{\Iii}}{{\Iiii}{\Ii}{\Iiii}{\Ii}}{{\Iii}{1}{\Iii}{1}}{{\Ii}{\Iiii}{\Ii}{\Iiii}}\nolinebreak
\ \ \makebox[0mm][l]{$\chi_2$:}\!\!\!
\pcSH{white}{{1}{1}{1}{1}}{{\Iii}{\Iii}{\Iii}{\Iii}}{{1}{1}{1}{1}}{{\Iii}{\Iii}{\Iii}{\Iii}}\nolinebreak
\ \ \makebox[0mm][l]{$\chi_{-1}$:}\!\!\!
\pcSH{white}{{1}{\Iii}{1}{\Iii}}{{\Ii}{\Iiii}{\Ii}{\Iiii}}{{\Iii}{1}{\Iii}{1}}{{\Iiii}{\Ii}{\Iiii}{\Ii}}

} % \def\picEEEE{

\def\picSH#1#2#3#4#5{\picSh{#2}{#3}{#4}{#5}{#1}}
\def\picSh#1#2#3#4#5{
\begin{tikzpicture}[
xscale=0.45,yscale=0.45,xslant=-0.0,
nn/.style={circle,fill=#5,draw=black,%thick, 
           inner sep=1pt}]
           \begin{scope}  
\clip [] (-0.45,-0.45) rectangle (3.45,3.45);
\draw[ystep=1,xstep=1,] (-4.9,-2.1) grid (5.4,3.9);
\draw[xslant=1,ystep=9,xstep=1,] (-3.4,-2.1) grid (6.4,3.9);
\end{scope}
\draw 
  (0,0) node [nn] {\phZ[\Aa#1]} +(1,0) node [nn] {\phZ[\Bb#1]} +(2,0) node [nn] {\phZ[\Cc#1]} +(3,0) node [nn] {\phZ[\Dd#1]}
++(0,1) node [nn] {\phZ[\Aa#2]} +(1,0) node [nn] {\phZ[\Bb#2]} +(2,0) node [nn] {\phZ[\Cc#2]} +(3,0) node [nn] {\phZ[\Dd#2]}
++(0,1) node [nn] {\phZ[\Aa#3]} +(1,0) node [nn] {\phZ[\Bb#3]} +(2,0) node [nn] {\phZ[\Cc#3]} +(3,0) node [nn] {\phZ[\Dd#3]}
++(0,1) node [nn] {\phZ[\Aa#4]} +(1,0) node [nn] {\phZ[\Bb#4]} +(2,0) node [nn] {\phZ[\Cc#4]} +(3,0) node [nn] {\phZ[\Dd#4]};
\end{tikzpicture}
}

\def\picZZZZ{
$$\begin{array}{r@{\ }rr@{\ }rr@{\ }rr@{\ }rl}
\raisebox{10mm}{$\chi_{30}:$}&
\picSH{gray!40!white}{{\Iiiii}{\Iiiii}{\Iiiii}{\Iiiii}}{{\Iiii}{\Iiii}{\Iiii}{\Iiii}}{{\Iii}{\Iii}{\Iii}{\Iii}}{{\Ii}{\Ii}{\Ii}{\Ii}}&
\raisebox{10mm}{$\chi_{31}:$}&
\picSH{gray!40!white}{{\Iiiii}{\Ii}{\Iii}{\Iiii}} {{\Iiii}{\Iiiii}{\Ii}{\Iii}} {{\Iii}{\Iiii}{\Iiiii}{\Ii}} {{\Ii}{\Iii}{\Iiii}{\Iiiii}}&
\raisebox{10mm}{$\chi_{32}:$}&
\picSH{white}{{\Iiiii}{\Iii}{\Iiiii}{\Iii}}{{\Iiii}{\Ii}{\Iiii}{\Ii}}{{\Iii}{\Iiiii}{\Iii}{\Iiiii}}{{\Ii}{\Iiii}{\Ii}{\Iiii}}&
\raisebox{10mm}{$\chi_{33}:$}&
\picSH{white}{{\Iiiii}{\Iiii}{\Iii}{\Ii}}{{\Iiii}{\Iii}{\Ii}{\Iiiii}}{{\Iii}{\Ii}{\Iiiii}{\Iiii}}{{\Ii}{\Iiiii}{\Iiii}{\Iii}}&
\\[1mm]
\noindent\mbox{}\hfill\mbox{}%
\raisebox{10mm}{$\chi_{20}:$}&
\picSH{white}{{\Iiiii}{\Iiiii}{\Iiiii}{\Iiiii}}{{\Iii}{\Iii}{\Iii}{\Iii}}{{\Iiiii}{\Iiiii}{\Iiiii}{\Iiiii}}{{\Iii}{\Iii}{\Iii}{\Iii}}&
\raisebox{10mm}{$\chi_{21}:$}&
\picSH{white}{{\Iiiii}{\Ii}{\Iii}{\Iiii}}{{\Iii}{\Iiii}{\Iiiii}{\Ii}}{{\Iiiii}{\Ii}{\Iii}{\Iiii}}{{\Iii}{\Iiii}{\Iiiii}{\Ii}}&
\raisebox{10mm}{$\chi_{22}:$}&
\picSH{white}{{\Iiiii}{\Iii}{\Iiiii}{\Iii}}{{\Iii}{\Iiiii}{\Iii}{\Iiiii}}{{\Iiiii}{\Iii}{\Iiiii}{\Iii}}{{\Iii}{\Iiiii}{\Iii}{\Iiiii}}&
\raisebox{10mm}{$\chi_{23}:$}&
\picSH{white}{{\Iiiii}{\Iiii}{\Iii}{\Ii}} {{\Iii}{\Ii}{\Iiiii}{\Iiii}} {{\Iiiii}{\Iiii}{\Iii}{\Ii}} {{\Iii}{\Ii}{\Iiiii}{\Iiii}}&
\\[1mm]
\noindent\mbox{}\hfill\mbox{}%
\raisebox{10mm}{$\chi_{10}:$}&
\picSH{gray!40!white}{{\Iiiii}{\Iiiii}{\Iiiii}{\Iiiii}}{{\Ii}{\Ii}{\Ii}{\Ii}}{{\Iii}{\Iii}{\Iii}{\Iii}}{{\Iiii}{\Iiii}{\Iiii}{\Iiii}}&
\raisebox{10mm}{$\chi_{11}:$}&
\picSH{white}{{\Iiiii}{\Ii}{\Iii}{\Iiii}} {{\Ii}{\Iii}{\Iiii}{\Iiiii}} {{\Iii}{\Iiii}{\Iiiii}{\Ii}} {{\Iiii}{\Iiiii}{\Ii}{\Iii}}&
\raisebox{10mm}{$\chi_{12}:$}&
\picSH{white}{{\Iiiii}{\Iii}{\Iiiii}{\Iii}}{{\Ii}{\Iiii}{\Ii}{\Iiii}}{{\Iii}{\Iiiii}{\Iii}{\Iiiii}}{{\Iiii}{\Ii}{\Iiii}{\Ii}}&
\raisebox{10mm}{$\chi_{13}:$}&
\picSH{gray!40!white}{{\Iiiii}{\Iiii}{\Iii}{\Ii}}{{\Ii}{\Iiiii}{\Iiii}{\Iii}}{{\Iii}{\Ii}{\Iiiii}{\Iiii}}{{\Iiii}{\Iii}{\Ii}{\Iiiii}}&
\\[1mm]
\noindent\mbox{}\hfill\mbox{}%
\raisebox{10mm}{$\chi_{00}:$}&
\picSH{gray}{{\Iiiii}{\Iiiii}{\Iiiii}{\Iiiii}}{{\Iiiii}{\Iiiii}{\Iiiii}{\Iiiii}}{{\Iiiii}{\Iiiii}{\Iiiii}{\Iiiii}}{{\Iiiii}{\Iiiii}{\Iiiii}{\Iiiii}}&
\raisebox{10mm}{$\chi_{01}:$}&
\picSH{gray!40!white}{{\Iiiii}{\Ii}{\Iii}{\Iiii}}{{\Iiiii}{\Ii}{\Iii}{\Iiii}}{{\Iiiii}{\Ii}{\Iii}{\Iiii}}{{\Iiiii}{\Ii}{\Iii}{\Iiii}}&
\raisebox{10mm}{$\chi_{02}:$}&
\picSH{white}{{\Iiiii}{\Iii}{\Iiiii}{\Iii}}{{\Iiiii}{\Iii}{\Iiiii}{\Iii}}{{\Iiiii}{\Iii}{\Iiiii}{\Iii}}{{\Iiiii}{\Iii}{\Iiiii}{\Iii}}&
\raisebox{10mm}{$\chi_{03}:$}&
\picSH{gray!40!white}{{\Iiiii}{\Iiii}{\Iii}{\Ii}} {{\Iiiii}{\Iiii}{\Iii}{\Ii}} {{\Iiiii}{\Iiii}{\Iii}{\Ii}} {{\Iiiii}{\Iiii}{\Iii}{\Ii}}&
\\[-1mm]
\end{array}
$$

} % \def\picZZZZ

\def\picE{
\begin{tikzpicture}[xslant=-0.577,yscale=.866,
scale=0.8,
nn/.style={circle,draw=black,thin, 
           inner sep=1.5pt}]
           \begin{scope} %xslant=0.577,
\clip [] (-0.45,-0.45) rectangle (3.45,3.45);
\draw[ystep=1,xstep=1, thin] (-4.9,-2.1) grid (5.4,3.9);
\draw[xslant=1,ystep=9,xstep=1, thin] (-3.4,-2.1) grid (6.4,3.9);
\end{scope}
\draw [dashed,thin] (-0.5,-0.5) rectangle (3.5,3.5);
\draw [dashed,thin,<->,rounded corners=3.5mm] (-0.5,2.7) -- (-1.2,2.7) -- (-1.2,3.8) -- (4.2,3.8) --(4.2,2.7) -- (3.5,2.7);
\draw [dashed,thin,<->,rounded corners=3.5mm] (-0.5,0.3) -- (-1.2,0.3) -- (-1.2,-0.8) -- (4.2,-0.8) --(4.2,0.3) -- (3.5,0.3);
\draw [dashed,thin,<->,rounded corners=3.5mm] (0.3,-0.5) -- (0.3,-1.2) -- (-0.8,-1.2) -- (-0.8,4.2) --(0.3,4.2) -- (0.3,3.5);
\draw [dashed,thin,<->,rounded corners=3.5mm] (2.7,-0.5) -- (2.7,-1.2) -- (3.8,-1.2) -- (3.8,4.2) --(2.7,4.2) -- (2.7,3.5);
\draw 
  (0,0) node [nn,fill=gray!60!white] {\phantom{0}\makebox[0pt][c]{$0$}\phantom{0}} 
  +(1,0) node [nn,fill=green!30!white] {\phantom{0}\makebox[0pt][c]{$1$}\phantom{0}} 
  +(2,0) node [nn,fill=white] {\phantom{0}\makebox[0pt][c]{$2$}\phantom{0}} 
  +(3,0) node [nn,fill=green!30!white] {\phantom{0}\makebox[0pt][c]{$-1$}\phantom{0}}
++(0,1) node [nn,fill=green!30!white] {\phantom{0}\makebox[0pt][c]{$\omega$}\phantom{0}} 
+(1,0) node [nn,fill=green!30!white] {\phantom{0}\makebox[0pt][c]{$-\bar\omega$}\phantom{0}} 
+(2,0) node [nn,fill=red!20!white] {\phantom{0}\makebox[0pt][c]{$\bar\omega\psi$}\phantom{0}} 
+(3,0) node [nn,fill=red!20!white] {\phantom{0}\makebox[0pt][c]{$\psi$}\phantom{0}}
++(0,1) node [nn,fill=white] {\phantom{0}\makebox[0pt][c]{$2\omega$}\phantom{0}} 
+(1,0) node [nn,fill=red!20!white] {\phantom{0}\makebox[0pt][c]{$-\!\omega\!\psi$}\phantom{0}} 
+(2,0) node [nn,fill=white] {\phantom{0}\makebox[0pt][c]{$2\bar\omega$}\phantom{0}} 
+(3,0) node [nn,fill=red!20!white] {\phantom{0}\makebox[0pt][c]{$\omega\psi$}\phantom{0}}
++(0,1) node [nn,fill=green!30!white] {\phantom{0}\makebox[0pt][c]{$-\omega$}\phantom{0}} 
+(1,0) node [nn,fill=red!20!white] {\phantom{0}\makebox[0pt][c]{$-\psi$}\phantom{0}} 
+(2,0) node [nn,fill=red!20!white] {\phantom{0}\makebox[0pt][c]{$-\!\bar\omega\!\psi$}\phantom{0}} 
+(3,0) node [nn,fill=green!30!white] {\phantom{0}\makebox[0pt][c]{$\omega$}\phantom{0}};
\end{tikzpicture}
}

\def\picZ{
\begin{tikzpicture}[
scale=0.6,
nn/.style={circle,fill=white,draw=black, 
           inner sep=0.3pt}]
           \begin{scope} %xslant=0.577,
\clip [] (-0.45,-0.45) rectangle (3.45,3.45);
\draw[ystep=1,xstep=1, thick] (-4.9,-2.1) grid (5.4,3.9);
\draw[xslant=1,ystep=9,xstep=1, thick] (-3.4,-2.1) grid (6.4,3.9);
\end{scope}
\draw [dashed,thin] (-0.5,-0.5) rectangle (3.5,3.5);
\draw [dashed,thin,<->,rounded corners=3.5mm] (-0.5,2.7) -- (-1.2,2.7) -- (-1.2,3.8) -- (4.2,3.8) --(4.2,2.7) -- (3.5,2.7);
\draw [dashed,thin,<->,rounded corners=3.5mm] (-0.5,0.3) -- (-1.2,0.3) -- (-1.2,-0.8) -- (4.2,-0.8) --(4.2,0.3) -- (3.5,0.3);
\draw [dashed,thin,<->,rounded corners=3.5mm] (0.3,-0.5) -- (0.3,-1.2) -- (-0.8,-1.2) -- (-0.8,4.2) --(0.3,4.2) -- (0.3,3.5);
\draw [dashed,thin,<->,rounded corners=3.5mm] (2.7,-0.5) -- (2.7,-1.2) -- (3.8,-1.2) -- (3.8,4.2) --(2.7,4.2) -- (2.7,3.5);
\draw 
  (0,0) node [nn,fill=gray!40!white] {$ 00$} +(1,0) node [nn,fill=green!20!white] {$ 10$} +(2,0) node [nn] {$ 20$} +(3,0) node [nn,fill=green!20!white] {$ 30$}
++(0,1) node [nn,fill=green!20!white] {$ 01$} +(1,0) node [nn,fill=green!20!white] {$ 11$} +(2,0) node [nn] {$ 21$} +(3,0) node [nn] {$ 31$}
++(0,1) node [nn] {$ 02$} +(1,0) node [nn] {$ 12$} +(2,0) node [nn] {$ 22$} +(3,0) node [nn] {$ 32$}
++(0,1) node [nn,fill=green!20!white] {$ 03$} +(1,0) node [nn] {$ 13$} +(2,0) node [nn] {$ 23$} +(3,0) node [nn,fill=green!20!white] {$ 33$};
\end{tikzpicture}
}

\def\picZZ{
\begin{tikzpicture}[
scale=0.6,
nn/.style={circle,fill=white,draw=black, 
           inner sep=0.3pt}]
           \begin{scope} %xslant=0.577,
\clip [] (-0.45,-0.45) rectangle (3.45,3.45);
\draw[ystep=1,xstep=1, thick] (-4.9,-2.1) grid (5.4,3.9);
\draw[xslant=-1,ystep=9,xstep=1, thick] (-3.4,-2.1) grid (6.4,3.9);
\end{scope}
\draw [dashed,thin] (-0.5,-0.5) rectangle (3.5,3.5);
\draw [dashed,thin,<->,rounded corners=3.5mm] (-0.5,2.7) -- (-1.2,2.7) -- (-1.2,3.8) -- (4.2,3.8) --(4.2,2.7) -- (3.5,2.7);
\draw [dashed,thin,<->,rounded corners=3.5mm] (-0.5,0.3) -- (-1.2,0.3) -- (-1.2,-0.8) -- (4.2,-0.8) --(4.2,0.3) -- (3.5,0.3);
\draw [dashed,thin,<->,rounded corners=3.5mm] (0.3,-0.5) -- (0.3,-1.2) -- (-0.8,-1.2) -- (-0.8,4.2) --(0.3,4.2) -- (0.3,3.5);
\draw [dashed,thin,<->,rounded corners=3.5mm] (2.7,-0.5) -- (2.7,-1.2) -- (3.8,-1.2) -- (3.8,4.2) --(2.7,4.2) -- (2.7,3.5);
\draw 
  (0,0) node [nn,fill=gray!40!white] {$ 00$} +(1,0) node [nn,fill=red!20!white] {$ 10$} +(2,0) node [nn] {$ 20$} +(3,0) node [nn,fill=red!20!white] {$ 30$}
++(0,1) node [nn,fill=red!20!white] {$ 01$} +(1,0) node [nn] {$ 11$} +(2,0) node [nn] {$ 21$} +(3,0) node [nn,fill=red!20!white] {$ 31$}
++(0,1) node [nn] {$ 02$} +(1,0) node [nn] {$ 12$} +(2,0) node [nn] {$ 22$} +(3,0) node [nn] {$ 32$}
++(0,1) node [nn,fill=red!20!white] {$ 03$} +(1,0) node [nn,fill=red!20!white] {$ 13$} +(2,0) node [nn] {$ 23$} +(3,0) node [nn] {$ 33$};
\end{tikzpicture}
}

\begin{document}
\title{On Dual Codes in the Doob Schemes} 
\author{%
   \IEEEauthorblockN{Denis~S.~Krotov%
}
   \IEEEauthorblockA{Sobolev Institute of Mathematics,
                     Novosibirsk 630090, Russia\\
                     Email: krotov@math.nsc.ru, dk@ieee.org}
}

\maketitle

\begin{abstract}\boldmath
The Doob scheme $D(m,n'+n'')$ is a metric association scheme defined 
on $\mathbb{E}_4^m \times \mathbb{F}_4^{n'}\times\mathbb{Z}_4^{n''}$, 
where $\mathbb{E}_4=\mathrm{GR}(4^2)$ or, alternatively, on
$\mathbb{Z}_4^{2m} \times \mathbb{Z}_2^{2n'} \times\mathbb{Z}_4^{n''}$.
We prove the MacWilliams identities connecting the weight distributions
of a linear or additive code and its dual.
In particular, for each case, we determine the dual scheme,
on the same set but with different metric, 
such that the weight distribution of an additive code $C$
in the Doob scheme $D(m,n'+n'')$ is related by the MacWilliams identities
with the weight distribution of the dual code $C^\perp$
in the dual scheme. We note that in the case of a linear code $C$ in 
$\mathbb{E}_4^m \times \mathbb{F}_4^{n'}$, the weight distributions
of $C$ and $C^\perp$ in the same scheme are also connected.
\end{abstract}

\def\Z{\mathbb Z}
\def\F{\mathbb F}
\def\E{\mathbb E}
\def\C{\mathbb C}
\def\R{\mathbb R}
\def\V{V_{m,n',n''}}
\def\inn#1#2{{\langle #1,#2 \rangle}}
\def\innb#1#2{\inn{\vec{#1}}{\vec{#2}}}
\newcommand\zinn[3][{}]{{[ #2,#3 ]_{#1}}}
\newcommand\zinnb[3][{}]{\zinn[#1]{\vec{#2}}{\vec{#3}}}
\def\vec#1{{\boldsymbol{#1}}}
\def\2{2\cdot }
\newcommand\Tr[1][{}]{\mathrm{Tr}_{#1}}
\newcommand\wt{\mathrm{wt}}
\newcommand\wa{\widetilde{\phantom{wa}}\!\!\!\!\!\!\!\mathrm{wt}}
\section{Introduction}
{T}{he} codes in Doob graphs are special cases of codes over Eisenstein--Jacobi integers, see, e.g., \cite{Huber94,MSBG:2008}, which can be used for the information transmission in the channels with two-dimensional or complex-valued modulation. The vertices of a Doob graph can be considered as words in the mixed alphabet consisting of the elements of the quotient (modulo $4$ and modulo $2$) rings of the ring of Eisenstein--Jacobi integers, see, e.g., \cite{Kro:perfect-doob}. In contrast to the cases considered in \cite{Huber94,MSBG:2008}, $4$ is not a prime number, and the quotient ring is not a field. This fact is not a problem from the point of view of the modern coding theory, which has a reach set of algebraic and combinatorial tools to deal with rings, see, e.g., \cite{SAS:codes&rings}; moreover, studying codes in the Doob graphs is additionally motivated by the application of association schemes in coding theory \cite{Delsarte:1973}: 
the algebraic parameters of the schemes associated with these graphs are the same as for the quaternary Hamming scheme (this fact can be also treated from the point of view of the corresponding distance-regular graphs).

In this correspondence, we establish the MacWilliams identities for additive and linear codes in Doob graphs. In particular, for each standard inner product, we determine the Delsarte dual to the Doob graph, and suggest inner products such that the Delsarte dual coincides with the original Doob graph. 
\section[Codes in E4m x F4n' x Z4n'']{Codes in $\mathbb{E}_4^m \times \mathbb{F}_4^{n'}\times\mathbb{Z}_4^{n''}$}

\subsection{Linear codes and duality}\label{s:lin}

\renewcommand\thefootnote{}
\footnotetext{This work was funded by the Russian Science Foundation under grant\linebreak 18-11-00136}%
Let 
$i$ be a square root of $-1$, let
$$ \omega:=\frac{-1+i\sqrt3}{2} $$
be a primitive cube root of $1$,
and for any complex number 
$x=a+bi\in \C$, $a,b\in \R$,
let $\bar x$ denote its conjugate, 
$\bar x:=a-bi$.
The numbers of form 
$$ a+b\omega, \quad a,b\in \Z$$
are known as the \emph{Eisenstein--Jacobi integers}. 
Endowed with the complex addition and multiplication,
they form a subring of $\C$, which will be denoted by $\E$.

By $\E_p$, $p\in \E$, we denote the factor ring $\E/p\E$;
we are interested in $\E_2$,
which is the finite field $\F_4$ of order $4$,
and $\E_4$, which is the Galois ring GR$(4^2)$.
The set of tuples from the Cartesian product 
$V_{m,n}:=\E_4^m \times \F_4^n$ of $m$ copies of $\E_4$ 
and $n$  copies of $\F_4$ with the component-wise addition
and multiplication by a constant (modulo $4$ in the first $m$ positions, 
modulo $2$ in the last $n$ positions) forms a module over $\E_4$.

In $\E_4$, we consider the set of powers of $-\omega$  and call 
it  $\Omega$:
$$\Omega := \{\pm 1,\pm \omega,\pm \bar\omega\}.$$
It forms a multiplicative subgroup of $\E_4$, and $\E_4$ is partitioned 
into the multiplicative cosets $0\Omega$, $\Omega$, $2\Omega$, and $\psi\Omega$, where
$$\psi:=\omega-1. $$

Any nonempty subset of $V_{m,n}$ closed under addition
and multiplication by a constant is called a \emph{linear code} or a 
\emph{linear $\F_4\E_4$ code}. 
So, the linear codes correspond to the submodules of $V_{m,n}$.

Any nonempty subset of $V_{m,n}$ closed under addition
is called an \emph{additive code} or an 
\emph{additive $\F_4\E_4$ code}. 

For $\vec x=(x_1, \ldots ,x_k)$ and $\vec y=(y_1, \ldots ,y_k)$ from $\E_4^k$ or $\F_4^k$, 
the standard inner product
$\innb{x}{y}$ is defined as usual:
\begin{equation}\label{eq:in1}
\innb{x}{y} := x_1y_1+  \ldots  +x_k y_k.
\end{equation}

For $\vec x=(\vec x^*,\vec x')$ and $\vec y=(\vec y^*,\vec y')$, where 
$\vec x^*,\vec y^*\in\E_4^m$, $m>0$, and $\vec x',\vec y'\in\F_4^n$, 
the inner product is defined as 
\begin{equation}\label{eq:in2}
\innb{x}{y} := \innb{x^*}{y^*}+\2\innb{x'}{y'},
\end{equation}
where $\2$ is the group homomorphism
$z+2\E \to 2z + 4\E$ from $\F_4^+$ to $\E_4^+$.

Two vectors $\vec x$ and $\vec y$ are said to be \emph{orthogonal}
if $\innb{x}{y}=0$.
For a linear code $C$ in $V_{m,n}$,
its \emph{dual},
denoted $C^\perp$, 
is defined as the set of all vectors from $V_{m,n}$
that are orthogonal to every vector from $C$.
The dual $C^\perp$
is also a linear code and one has $C^{\perp\perp}=C$ and $|C|\cdot |C^\perp|=|V_{m,n}|$.
%, where ${<}C{>}$ is the \emph{linear span} of $C$
%(if $C$ is a linear code, then ${<}C{>}=C$).

\subsection{Additive codes and duality}\label{s:add}
For codes over finite fields,
there is a standard technique to extend the concept of duality from the linear codes
to the additive codes, below, we describe the approach in application to the ring $\E_4$.
The module $V_{m,n}$ over $\E_4$ 
can also be viewed as a module over the subring $\Z_4$ of $\E_4$. 
Given an inner product $\inn{\cdot}{\cdot}$ and a surjective
$\Z_4$-linear map $L$ from $\E_4$ onto $\Z_4$, 
we define the associated $\Z_4$-inner product $\zinn[L]{\cdot}{\cdot}$ as
$$ \zinn[L]{\cdot}{\cdot} := L(\inn{\cdot}{\cdot}). $$
Two vectors $\vec x$ and $\vec y$ are said to be \emph{$L$-orthogonal}
if $\zinnb[L]{x}{y}=0$.
Given an additive code $C$ in $V_{m,n}$, 
its \emph{$L$-dual} 
is defined as the set of all vectors from $V_{m,n}$
that are $L$-orthogonal to every vector from $C$.
The following known and easy fact shows that the $L$-duality
is an extension of the duality to the additive codes.
\begin{proposition}\label{p:ad}
 For every surjective
$\Z_4$-linear map $L$ from $\E_4$ onto $\Z_4$,
two linear codes are {$L$-dual} if and only if they are  {dual}.
\end{proposition}
\begin{IEEEproof}
Let $C$ be a linear code. Obviously, the orthogonality of two vectors implies their 
$L$-orthogonality. 
So, any vector from $C^\perp$ also belongs to the $L$-dual of $C$.
Let us consider a vector $\vec v$ from the $L$-dual. 
By the definition,
$$ L(\innb{v}{x})=0\qquad\mbox{for all } \vec x\in C.$$
Since $C$ is linear, 
\begin{equation}\label{eq:072} L(a\innb{v}{x})=L(\inn{\vec v}{a \vec x})=0\quad\mbox{for all } \vec x\in C,\ a\in\E_4.\end{equation}
Seeking a contradiction, assume that $b:=\innb{v}{x}\ne 0$ for some $\vec x$ from $C$.
Take $d\in \E_4$ such that $L(d)=1$.
There should be $a$ such that $ab=d$ or $ab=2d$. 
For this $a$, we have $L(a\innb{v}{x})\in\{1,2\}$,
contradicting \eqref{eq:072}.
Hence, for all $\vec x$ from $C$ we have $\innb{v}{x}= 0$, and $\vec v$ belongs to $C^\perp$.
\end{IEEEproof}

A standard linear map from $\E_4$ to $\Z_4$ (as well as from a finite field to its prime-order subfield) used for different purposed is the \emph{trace} $\Tr$:
$$ \Tr(z) := z + \bar z. $$
(In general, for an arbitrary Galois ring, $\Tr(z):=\sum\sigma(z)$, where the sum is over all automorphisms $\sigma$ of the ring.)
Below, for brevity, the $\Tr$-duality will be referred to as duality (by Proposition~\ref{p:ad}, 
it is in agree with the duality of linear codes),
and the $\Tr$-inner product $\zinn[\Tr]{\cdot}{\cdot}$ will be denoted as $\zinn{\cdot}{\cdot}$.
The notation $C^\perp$ is also extended to the additive codes $C$.

\subsection{Extending the concept of additive codes}\label{s:ext}

We can further extend the definition of the additive codes, the concept of duality,
and the corresponding notation
to the codes in 
$$V_{m,n',n''}:=\E_4^m \times \F_4^{n'} \times \Z_4^{n''}$$
by defining the following $Z_4$-inner product:
$$ \zinnb{x}{y}:=\Tr\innb{x^*}{y^*}+ 2\cdot\Tr\innb{x'}{y'} + \innb{x''}{y''}.$$
Here and below, for any vector $\vec z$ from $V_{m,n',n''}$, 
the notations $\vec z^*$, $\vec z'$, and $\vec z''$ denote the $\E_4$-, $\F_4$-,
and $\Z_4$-parts of $\vec z$, respectively; $z^*_i$, $z'_i$, and $z''_i$ refer to the $i$-th
coordinate of $\vec z^*$, $\vec z'$, or $\vec z''$, respectively. I.e., 
$$ \vec z = (\vec z^*,\vec z',\vec z'') = ( z^*_1, \ldots ,z^*_m, z'_1, \ldots ,z'_{n'}, z''_1, \ldots ,z''_{n''}). $$

The motivation for this extension lies in the extended possibility 
to construct codes with different parameters in the same metric space 
(as we will see in the next subsection, 
the coordinates from the last two groups are metrically equivalent in the considered scheme). 
Examples of exploiting this possibility can be found in the construction of additive 
$1$-perfect codes \cite{SHK:additive}.

\subsection{The Doob metric}\label{s:met}
We define two weight function on $V_{m,n',n''}$:
\begin{IEEEeqnarray*}{rCl}
 \wt(\vec x)&:=&\sum_{j=1}^{m} \wt(x^*_{j})+\sum_{j=1}^{n'} \wt(x'_{j})
+\sum_{j=1}^{n''} \wt(x''_{j}),\\
 \wa(\vec x)&:=&\sum_{j=1}^{m} \wa(x^*_{j})+\sum_{j=1}^{n'} \wt(x'_{j})
+\sum_{j=1}^{n''} \wt(x''_{j}),
\end{IEEEeqnarray*}
called the \emph{weight} and the \emph{coweight} of $\vec x$, where
$$
\wt(x^*_{j})=
\begin{cases}
0 & \mbox{if }x^*_{j}=0 \cr
1 & \mbox{if }x^*_{j}\in \Omega \cr
2 & \mbox{otherwise},
\end{cases}
\quad
\wa(x^*_{j})=
\begin{cases}
0 & \mbox{if }x^*_{j}=0 \cr
1 & \mbox{if }x^*_{j}\in \psi\Omega \cr
2 & \mbox{otherwise},
\end{cases}
$$
$$
\wt(x'_{j})=
\begin{cases}
0 & \mbox{if }x'_{j}=0 \cr
1 & \mbox{otherwise},
\end{cases}
\quad
\wt(x''_{j})=
\begin{cases}
0 & \mbox{if }x''_{j}=0 \cr
1 & \mbox{otherwise}.
\end{cases}
$$
Naturally, the weight function $\wt$ defines 
the distance
\begin{equation}\label{eq:d}
d(\vec x,\vec y):=\wt(\vec y-\vec x)
\end{equation}
and the distance-$1$ graph $D(m,n'+n'')$ on $V_{m,n',n''}$,
where two vectors are adjacent if and only if they are at distance $1$.
If $m>0$, then $D(m,n)$ is called a \emph{Doob graph} 
(the case $m=0$ corresponds to the quaternary Hamming graph $H(n,4)$).
The graph $D(m,n'+n'')$ is the Cartesian product of $m$ copies the \emph{Shrikhande graph}
$D(1,0)$
$$\picE
$$
and $n'+n''$ copies of the complete graph of order $4$
(with the vertex set $\F_4$ or $\Z_4$).

It is a distance-regular graph, 
and the corresponding metric induces an association scheme, 
a \emph{Doob scheme}.
From the metrical point of view, 
there is no difference between the $n=n'+n''$ last coordinates;
there are isometries of the metric space 
that permute these coordinates in an arbitrary manner.
The difference between these two groups 
becomes important only when we consider the additive structure, 
for example, when constructing additive codes.

In a similar way, $\wa$ defines another graph on $V_{m,n',n''}$, say 
$\widetilde D(m,n'+n'')$. It is isomorphic to $D(m,n'+n'')$ 
(an isomorphism is given by multiplying the first $m$ coordinates of the vectors by $\psi$).
From the theory below, one can see that the graphs $D(m,n'+n'')$ and $\widetilde D(m,n'+n'')$
are Delsarte dual to each other.
 \begin{figure*}[!t]
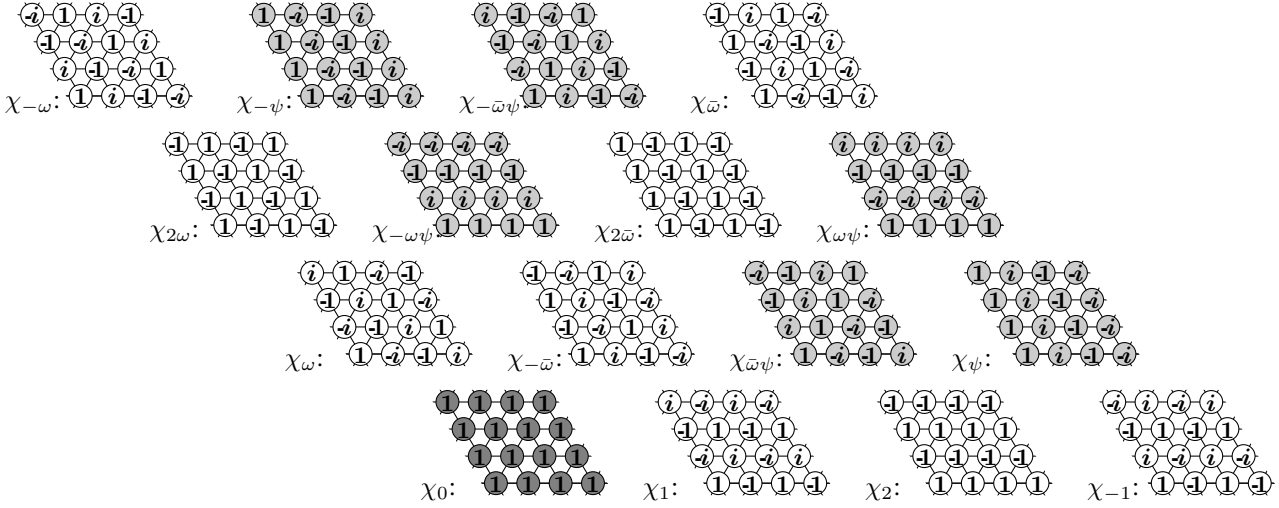

   \picEEEE
   \caption{Table of characters $\chi_v(u):=i^{\Tr(uv)}$, $u,v\in \E_4$.}
   \label{f:1}
 \end{figure*}

\section{MacWilliams identities}\label{s:MW}
For any complex-valued function $f$ on $V_{m,n',n''}$,
define its \emph{Fourier transform} $\widehat f:  V_{m,n',n''} \to \C$:
$$
\widehat f(\vec u):=\sum_{\vec v\in V_{m,n',n''}}i^{\zinnb{u}{v}} f(\vec v).
$$

\begin{lemma}\label{l:four}
Let ${C}$ be an additive code in 
$V_{m,n',n''}$,
% $\E_4^m\times \F_4^{n'}\times \Z_4^{n''}$, 
and let ${C}^\bot$ be its dual. 
Then for every complex-valued function $f$ on
$V_{m,n',n''}$,
% $\E_4^m\times \F_4^{n'}\times \Z_4^{n''}$,
\begin{equation}\label{eq:CC}
\sum\limits_{{\vec z}\in {C}^\bot}f(\vec z)=\frac{1}{|{C}|}\sum\limits_{{\vec u}\in {C}}\widehat{f}({\vec u}). 
\end{equation}
\end{lemma}
 
Consider the function
$$ 
f(\vec v):=
\prod_{j=1}^m X_{v^*_j} 
\prod_{j=1}^{n'} Y_{v'_j} 
\prod_{j=1}^{n''} Z_{v''_j}
$$
which also depends on the formal variables 
$X_a$, $a\in \E_4$, 
$Y_b$, $b\in \F_4$,  
$Z_0$, $Z_1$, $Z_2$, $Z_3$. 
The sum of $f$ over some code $C$ will be denoted by 
$$W_C(X_a, a\in \E_4; Y_b, b\in \F_4; Z_c, c\in \Z_4),\ \mbox{or } W_C(\vec X; \vec Y; \vec Z),$$ 
for short,
and called 
the \emph{complete weight enumerator} of $C$. The 
\emph{weight} and \emph{coweight enumerator}s of $C$
\begin{IEEEeqnarray}{rCl}\label{eq:WN}
 W_C({A},{B})&:=&\sum_{\vec z\in C}{A}^{N-\wt(\vec z)}{B}^{\wt(\vec z)}\quad\mbox{and}\\\nonumber
\widetilde W_C(\mathcal{A},\mathcal{B})&:=&\sum_{\vec z\in C}\mathcal{A}^{N-\wa(\vec z)}\mathcal{B}^{\wa(\vec z)},\quad N:=2m+n'+n'',
\end{IEEEeqnarray}
 respectively, are obtained from $W_C(\vec X; \vec Y; \vec Z)$ by identifying 
\begin{IEEEeqnarray}{l}\label{eq:id}
 X_a={A}^{2-\wt(a)}{B}^{\wt(a)}=\mathcal{A}^{2-\wa(a)}\mathcal{B}^{\wa(a)},\\
Y_0{=}Z_0={A}=\mathcal{A},\quad Y_{1}{=}Y_{\omega}{=}Y_{\bar\omega} {=} Z_1{=}Z_2{=}Z_3={B}=\mathcal{B},\nonumber
\end{IEEEeqnarray}
From the definitions of $f$ and the Fourier transform, we find
\begin{IEEEeqnarray*}{l}
\widehat f(\vec u)=\sum_{v\in \V} 
%i^{\Tr\innb{u^*}{v^*}+2\cdot \Tr\innb{u'}{v'}+\innb{u''}{v''}}
i^{\Tr\zinnb{u}{v}}
\prod_{j=1}^m X_{v^*_j} 
\prod_{j=1}^{n'} Y_{v'_j} 
\prod_{j=1}^{n''} Z_{v''_j}
\\ =
\prod_{j=1}^m    \sum_{v\in \E_4} i^{\Tr(u^*_jv)} X_v  \cdot
\prod_{j=1}^{n'}   \sum_{v\in \F_4} i^{2\cdot\Tr(u'_jv)} Y_v  \cdot  
\prod_{j=1}^{n''}    \sum_{v\in \Z_4} i^{u''_jv} Z_v.
\end{IEEEeqnarray*}
After expressing \eqref{eq:CC}, we get
\begin{IEEEeqnarray}{l} \label{eq:3905}
W_{C^\perp}(\vec X; \vec Y; \vec Z)\\
=\frac{1}{|C|}W_C 
\left( 
\sum_{v\in \E_4} i^{\Tr(a v)} X_v ;
\sum_{v\in \F_4} (-1)^{\Tr(b v)} Y_v;
\sum_{v\in \Z_4} i^{c v} Z_v
\right)\nonumber
\end{IEEEeqnarray}
We now identify the variables to get $W_{C^\perp}({A},{B})$ in the left side of \eqref{eq:3905}.
Straightforward calculations using the character table (Fig.~\ref{f:1}) show that 
\begin{IEEEeqnarray}{c}\label{eq:fromFig}
 \sum_{v\in \E_4} i^{\Tr(a v)} V^{\wt(v)} = ({A}+3{B})^{2-\wa(a)}({A}-{B})^{\wa(a)}.
\end{IEEEeqnarray}
Also, directly,
\begin{IEEEeqnarray*}{rCl}
\IEEEeqnarraymulticol{3}{l}{\sum_{v\in \F_4} (-1)^{\Tr(b v)} {A}^{1-\wt(v)}{B}^{\wt(v)} }
\\[-3mm]
\qquad\qquad\qquad\qquad\qquad  &= & ({A}{+}3{B})^{1-\wt(b)}({A}{-}{B})^{\wt(b)},\\[2mm]
\sum_{v\in \Z_4} i^{c v} {A}^{1-\wt(v)}{B}^{\wt(v)} & = & ({A}{+}3{B})^{1-\wt(c)}({A}{-}{B})^{\wt(c)}.
\end{IEEEeqnarray*}
Now, one can see that identifying the variables as in \eqref{eq:id} results in the first formula 
in the following theorem; the second formula is similar.
\begin{theorem}\label{th:E4additive}
 Let $C$ be an additive code in $\E_4^m\times \F_4^{n'} \times \Z_4^{n''}$.
 The Doob weight and coweight enumerators \eqref{eq:WN} 
 of $C$ and its dual $C^\perp$ are connected by the MacWilliams identities
\begin{IEEEeqnarray}{rCl} \label{eq:MWa1}
 W_{C^\perp}({A},{B}) &=& \frac{1}{|C|} \widetilde W_{C}({A}+{B},{A}-3{B}),\\
 \label{eq:MWa2}
 W_{C}({A},{B}) &=& \frac{1}{|C^\perp|} \widetilde W_{C^\perp}({A}+{B},{A}-3{B}).
\end{IEEEeqnarray}
\end{theorem}
Finally we note that in the case $n''=0$, 
each vector $\vec x$ from $\V$ satisfies 
$\wt(\vec x)=\wa(\psi\vec x)$. 
On the other hand, for a linear code $C$, $\vec x\in C$ is equivalent to $\psi\vec x\in C$.
Hence, $W_{C}({A},{B})=\widetilde W_{C}({A},{B})$ for linear codes, 
which is not true for additive codes in general. We conclude:
\begin{theorem}\label{th:E4lin}
 Let $C$ be a linear code in $\E_4^m\times \F_4^{n'}$.
 The Doob weight enumerators \eqref{eq:WN} of $C$ and its dual $C^\perp$ 
 satisfy
\begin{IEEEeqnarray}{rCl} \label{eq:MWl}
 W_{C^\perp}({A},{B}) &=& \frac{1}{|C|} W_{C}({A}+{B},{A}-3{B}).
\end{IEEEeqnarray}
\end{theorem}

For additive codes in general, the claim of Theorem~\ref{th:E4lin} is not always true.
For example, the additive codes $C=\{0,2,2\omega,2\bar\omega\}$ and $D=\{0,\psi,2\psi,-\psi\}$
have the same weight enumerator ${A}^2+3{B}^2$, but their duals 
$C^\perp=\{0,2,2\omega,2\bar\omega\}$ 
and 
$D^\perp=\{0,\omega,2\omega,-\omega\}$ 
have different weight enumerators ${A}^2+3{B}^2$ and
${A}^2+2{A} {B}+{B}^2$ (but the same coweight enumerator $\mathcal{A}^2+3\mathcal{B}^2$).
However,
we can modify the inner product, replacing $\Tr$ by another function, to make the weight distributions
of an additive code $C$ and its dual (in the sense of the modified inner product) be connected
my the MacWilliams identity.

\begin{theorem}\label{th:E4alt}
 Let $C$ be an additive code in $\E_4^m\times \F_4^{n'} \times \Z_4^{n''}$. 
 If $C^\perp$ is the dual of $C$ in the sence of the $Z_4$-inner product 
 $$\zinnb[(\psi)]{x}{y}:=\Tr(\psi\innb{x^*}{y^*})+\Tr\innb{x'}{y'}+\innb{x''}{y''},$$
 then the Doob weight enumerators \eqref{eq:WN} of $C$ and $C^\perp$ 
 satisfy the MacWilliams identity \eqref{eq:MWl}.
\end{theorem}

\begin{remark}
 In the theory of self-dual and quantum codes, 
 an important role is played by the Hermitian inner product defined as
 $$ \innb{x}{y}_{\mathrm{Herm}}:=x_1\bar y_1+\ldots +x_k\bar y_k$$
 for $\vec x=(x_1, \ldots ,x_k)$ and $\vec y=(y_1, \ldots ,y_k)$ from $\E_4^k$ or $\F_4^k$ (or, in general, for any other ring admitting an automorphisms $y\to\bar y$ of order $2$), see e.g. \cite{CRSS:quantum}.
 We note that the theory above remains valid if we replace the standard inner product 
 $\inn{\cdot}{\cdot}$ \eqref{eq:in1}
 by $\inn{\cdot}{\cdot}_{\mathrm{Herm}}$. 
 Indeed, obviously, the conjugancy does not change the weight; so, the dual code $C^\perp$
 and the Hermitian dual $\overline {C}\vphantom{C}^\perp$ have the same weight (or coweight) distribution.
\end{remark}

%====================================================
%====================================================
%====================================================

\section[Codes in Z42m x Z22n' x Z4n'']{Codes in $\Z_4^{2m} \times \Z_2^{2n'}\times\Z_4^{n''}$}

 \begin{figure*}[!t]
 \mbox{}\hfill \picZZZZ \hfill\mbox{}
   \caption{Table of characters $\chi_v(u):=i^{\inn{u}{v}}$, $u,v\in \Z_4^2$.}
   \label{f:2}
 \end{figure*}
 
As was mentioned above,
${\E}_4^m \times {\F}_4^{n'}\times{\Z}_4^{n''}$ can be considered 
as a $Z_4$-module, and in this sense it is isomorphic to the structure 
${\Z}_4^{2m} \times {\Z}_2^{2n'}\times{\Z}_4^{n''}$,
consisting of $(2m+2n'+n'')$-tuples with the first $2m$
and the last $n''$ elements from $\Z_4$ and the ``middle'' $2n'$ elements from $\Z_2$.
From this point of view, one can find convenient to consider the Doob scheme
directly on ${\Z}_4^{2m} \times {\Z}_2^{2n'}\times{\Z}_4^{n''}$
and consider the duality in a usual manner, as for $\Z_2\Z_4$-codes (see e.g. \cite{BFPRV2010}).

For convenience, we treat the elements of 
$({\Z}_4^2)^{m} \times (\Z_2^2)^{n'}\times{\Z}_4^{n''}$ as $(m+n'+n'')$-tuples,
where the first $m$ elements are pairs (we write like $a{:}b$, omitting ``{:}'' between numbers, e.g. $21=2{:}1$) from $\Z_4^2$,
 the next $n'$ elements are pairs from $\Z_2^2$,
 and the last $n''$ elements are from $\Z_4$.
 For vectors
 \begin{IEEEeqnarray*}{rCl}
 \vec x&:=&(x_1{:}y_1, \ldots ,x_m{:} y_m, x'_1{:} y'_1, \ldots ,x'_{n'} {:}y'_{n'}, x''_1, \ldots , x''_{n''}),
\\
\vec u&:=&(u_1{:}v_1, \ldots ,u_m {:}v_m, u'_1{:} v'_1, \ldots ,u'_{n'}{:} v'_{n'}, u''_1, \ldots , u''_{n''}),
 \end{IEEEeqnarray*}
the \emph{inner product $\innb{u}{x}$} is defined as
\begin{IEEEeqnarray}{r}
\IEEEeqnarraymulticol{1}{l}{\innb{u}{x}:=} \label{eq:innZ}
\\ \nonumber
\quad\quad\sum_{j=1}^m (u_jx_j{+}v_jy_j)
+2\cdot \sum_{j=1}^{n'}(u'_jx'_j{+}v'_jy'_j)
+ \sum_{j=1}^{n''}u''_jx''_j;
\end{IEEEeqnarray}
the \emph{weight} and \emph{coweight} are defined as
 \begin{IEEEeqnarray*}{rCl}
 \wt(\vec x)&:=&\sum_{j=1}^m \wt(x_j{:}y_j)
+\sum_{j=1}^{n'}\wt(x'_j{:}y'_j)
+\sum_{j=1}^{n''}\wt(x''_j),\\
 \wa(\vec x)&:=&\sum_{j=1}^m \wa(x_j{:}y_j)
+\sum_{j=1}^{n'}\wt(x'_j{:}y'_j)
+\sum_{j=1}^{n''}\wt(x''_j),
 \end{IEEEeqnarray*}
 where $\wt$ on $\Z_4^2$ is defined by $\wt(00)=0$,
 $\wt(a{:}b)=1$ for $a{:}b\in\{01,03,10,30,11,33\}$, 
 $\wt(a{:}b)=2$ in the other cases; 
 similarly,
 $\wa$ on $\Z_4^2$ is defined by $\wa(00):=0$,
 $\wa(a{:}b):=1$ for $a{:}b\in\{01,03,10,30,13,31\}$, 
 $\wa(a{:}b):=2$ in the other cases;
 on $\Z_2^2$ or $Z_4$ (in the last two groups of coordinates), 
 $\wt(00)=\wt(0):=0$ and 
 $\wt(01)=\wt(10)=\wt(11)=\wt(1)=\wt(2)=\wt(3):=1$.
As in the previous section, the corresponding metrics and graphs are defined.
The Shrikhande graph $D(1,0)$ and its dual $\widetilde D(1,0)$, with respect to the considered inner product, are as in the following picture:
$$
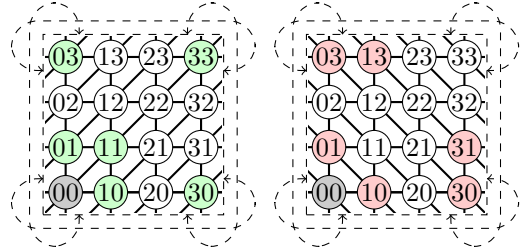
\picZ\ \  \picZZ
$$
In the notation presented in the current section,
the weight and coweight enumerators are defined as before \eqref{eq:WN}.
Moreover, the MacWilliams identities are also written as previously \eqref{eq:MWa1},
\eqref{eq:MWa2}, in the new notation.

\begin{theorem}\label{th:Z4additive}
 Let $C$ be an additive code in $(\Z_4^2)^m\times (\Z_2^2)_4^{n'} \times \Z_4^{n''}$ with the weight (coweight) functions \eqref{eq:innZ}.
 The Doob weight and coweight enumerators \eqref{eq:WN} of $C$ and its dual $C^\perp$ are connected by the MacWilliams identities
\begin{IEEEeqnarray*}{rCl}
 W_{C^\perp}({A},{B}) &=& \frac{1}{|C|} \widetilde W_{C}({A}+{B},{A}-3{B}),\\
 W_{C}({A},{B}) &=& \frac{1}{|C^\perp|} \widetilde W_{C^\perp}({A}+{B},{A}-3{B}).
\end{IEEEeqnarray*}
\end{theorem}

The technique to prove Theorem~\ref{th:Z4additive} is similar to that in the previous section. 
To check the identity
\begin{IEEEeqnarray*}{c}
 \sum_{v\in \Z_4^2} i^{\inn{a}{v}} V^{\wt(v)} = ({A}+3{B})^{2-\wa(a)}({A}-{B})^{\wa(a)},
\end{IEEEeqnarray*}
an analog of \eqref{eq:fromFig}, the character table (Fig.~\ref{f:2}) is useful.

Again, to make the weight distribution of an additive code and its dual connected in the same metric,
we can modify the duality by modifying the inner product.

\begin{theorem}\label{th:Z4alt}
 Let $C$ be an additive code in $(\Z_4^2)^m\times (\Z_2^2)^{n'} \times \Z_4^{n''}$. 
 If $C^\perp$ is the dual of $C$ in the sense of the modified inner product 
 \begin{IEEEeqnarray*}{c}
\innb{u}{x}_-:=
\sum_{j=1}^m (u_jx_j{\boldsymbol{-}}v_jy_j)
+2{\cdot}\!\! \sum_{j=1}^{n'}(u'_jx'_j{+}v'_jy'_j)
+ \sum_{j=1}^{n''}u''_jx''_j
\end{IEEEeqnarray*} 
% \begin{IEEEeqnarray*}{r}
% \IEEEeqnarraymulticol{1}{l}{\innb{u}{x}_-:=}
% \\ \nonumber
% \quad\quad\sum_{j=1}^m (u_jx_j{\boldsymbol{-}}v_jy_j)
% +2\cdot \sum_{j=1}^{n'}(u'_jx'_j{+}v'_jy'_j)
% + \sum_{j=1}^{n''}u''_jx''_j
% \end{IEEEeqnarray*}
(instead of \eqref{eq:innZ}),
then the Doob weight enumerators \eqref{eq:WN} of $C$ and $C^\perp$ 
 satisfy
 \begin{IEEEeqnarray*}{rCl}
  W_{C^\perp}({A},{B}) &=& \frac{1}{|C|} W_{C}({A}+{B},{A}-3{B}).
\end{IEEEeqnarray*}
\end{theorem}

\section{Conclusions}

\textit{1.} We conclude that to connect the weight distributions of an additive code and its dual in a Doob scheme by the MacWilliams
identities, one should consider their weights in different (but isomorphic) Doob metrics. An alternative approach
is to consider the weights in the same metric, but to treat duality in a modified way, with a special inner product, 
see Theorems~\ref{th:E4alt} and~\ref{th:Z4alt}. 
The last way can be recommended if a code is defined by a check matrix
which is treated as a generator matrix of the dual code.

\textit{2.}  It \cite{KoolMun:2000}, \cite{Kro:perfect-doob}, and \cite{SHK:additive}, 
classes of linear and additive $1$-perfect codes in Doob graphs were constructed.
By Theorem~\ref{th:E4lin}, the code generated by a check matrix 
of a linear $1$-perfect code in $D(m,n)$ has the parameters 
of the code dual to the linear $1$-perfect quaternary Hamming code of length $2m+n$ 
(the dual code is a simplex code, or a tight $2$-design, see e.g. \cite{KoolMun:2000}).
According to Theorem~\ref{th:Z4alt}, to get a generator matrix 
of a tight $2$-design from a check matrix of an additive
(over $\Z_4$) $1$-perfect code in a Doob graph $D(m,n'+n'')$, 
one should multiply by $-1$ the first $m$ columns of the matrix with even numbers.

\section*{Acknowledgment}
The author is grateful to Minjia Shi, Jack Koolen, Vladimir Potapov, and Patrick Sol\'e for useful discussions.
\bigskip

This work was funded by the Russian Science Foundation under grant 18-11-00136.

\bigskip\bigskip\bigskip\bigskip

% \bibliographystyle{IEEEtranS}
% \bibliography{../../k}
% \end{document}

% Generated by IEEEtranS.bst, version: 1.14 (2015/08/26)
\providecommand\href[2]{#2} \providecommand\url[1]{\href{#1}{#1}}
  \def\DOI#1{{\small {DOI}:
  \href{http://dx.doi.org/#1}{#1}}}\def\DOIURL#1#2{{\small{DOI}:
  \href{http://dx.doi.org/#2}{#1}}}

\end{document}